\documentclass[12pt]{article}
\usepackage{graphicx}
\begin{document}

\centerline{\bf Altruism and Antagonistic Pleiotropy in Penna Ageing Model}
\bigskip

\centerline{Stanis{\l}aw Cebrat$^1$ and Dietrich Stauffer$^{1,2}$}
\bigskip

\noindent
$^1$ Institute of Genetics and Microbiology,
University of Wroc{\l}aw,\\ ul. Przybyszewskiego 63/77, PL-54148 Wroc{\l}aw, Poland \\

\noindent
$^2$ Institute for Theoretical Physics, Cologne University, D-50923 K\"oln, Euroland \\

\noindent
email: cebrat@microb.uni.wroc.pl, stauffer@thp.uni-koeln.de\\

\begin{abstract}
The Penna ageing model is based on mutation accumulation theory. We show that 
it also allows for self-organization of antagonistic pleiotropy which helps 
at young age at the expense of old age. This can be interpreted as emergence
of altruism.

\end{abstract}


\section{Introduction}

How can altruism arise through Darwinian evolution of the fittest? Why does
a bee kill itself but saves many other bees by stinging a large predator? This 
question worried biologists as well as physicists since a long time 
\cite{napoli}.  Here we check if the Penna ageing model, suitably modified, 
allows for self-organization of altruism in the sense that the old make 
sacrifices for the young, or the young for the old. 

The Penna bitstring model \cite{penna,book,vancouver} is based on Medawar's
half-century old mutation-accumulation hypothesis: Dangerous mutations killing
an organism in young age will not be transmitted to future generations, while 
those affecting old age will be given on to the offspring. This model gave good
agreement with the exponential increase of mortality at middle age (Gompertz 
law). An alternative theory is antagonistic pleiotropy \cite{partridge}, where 
the same mutation has beneficial effects in young age but detrimental effects in
old age, like enhanced calcium intake producing bones for the young and 
arteriosclerosis for the old. Antagonistic pleiotropy was already combined with
the Penna ageing model for different purposes \cite{suzana}; here we use it
to explain altruism. A somewhat similar theory \cite{klotz} explained the shorter
life expectancy of men compared with women by testosterone, which makes men
more aggressive to protect their family; we like this heroic interpretation
better than explanations due to too much alcohol and steaks. In the next section
we define our models, followed by a section of results. 
 
\begin{figure}[hbt]
\begin{center}
\includegraphics[angle=-90,scale=0.5]{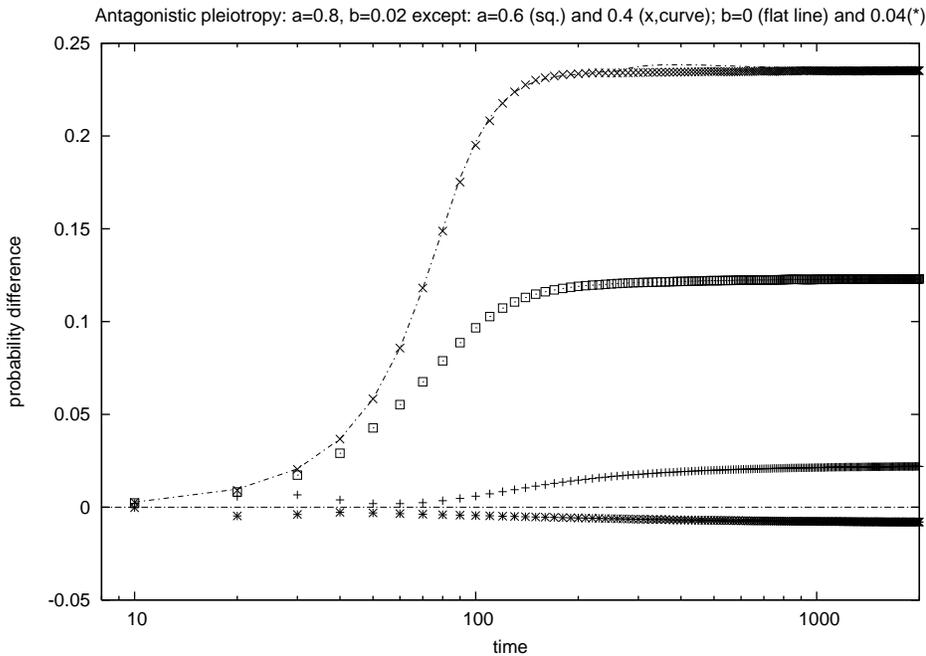}
\end{center}
\caption{$\Delta$ versus $t$ for various choices of the probability parameters
$a$ and $b$; $y=6, \; o=10$, final population several millions. The line gives
a ten times higher population than the x and was also (not shown) continued to 
$t = 20,000$. Positive $\Delta$ means the old help the young.
}
\end{figure}

\begin{figure}[hbt]
\begin{center}
\includegraphics[angle=-90,scale=0.5]{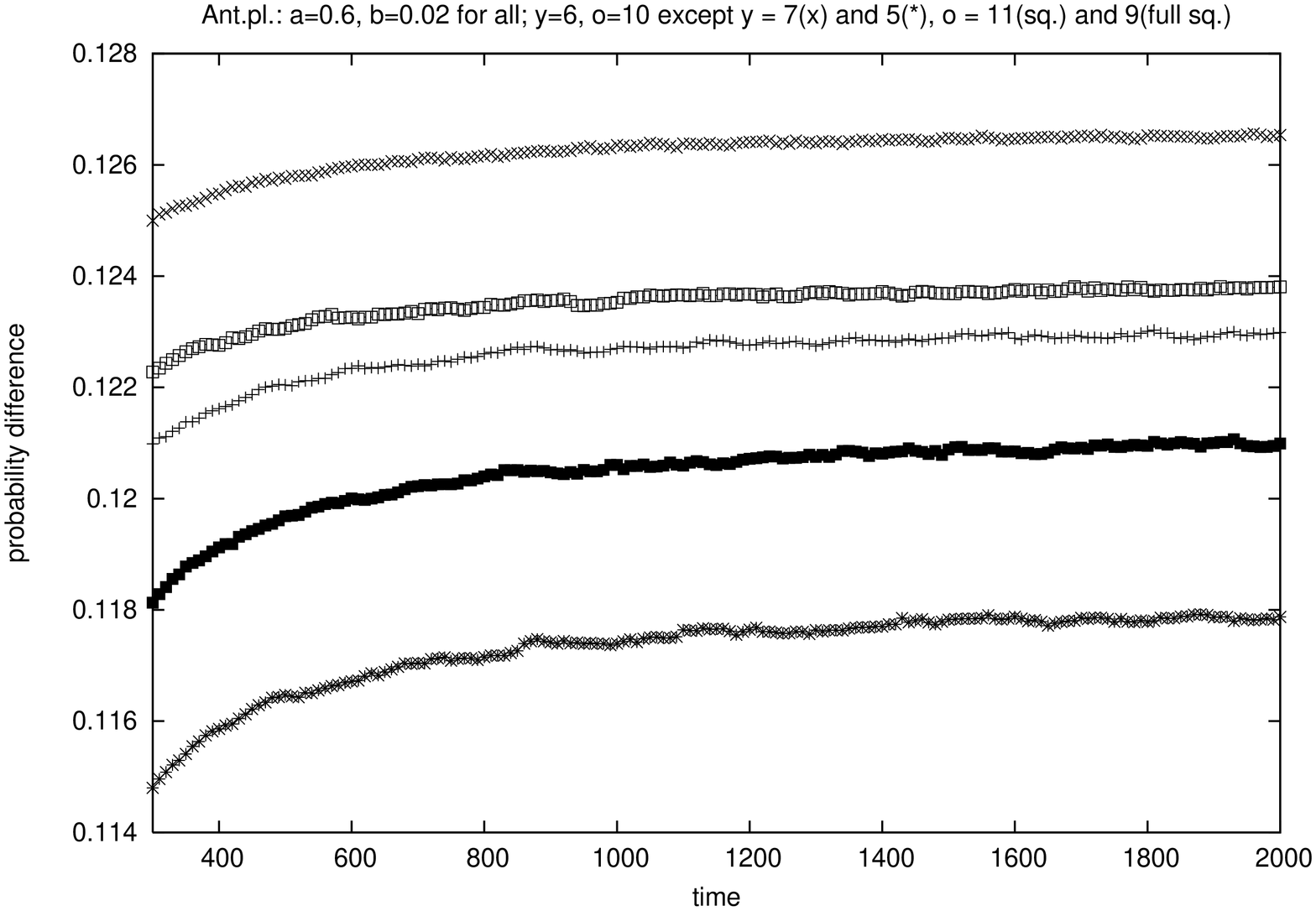}
\end{center}
\caption{$\Delta$ versus $t$ for various choices of the age threshold $y$ for 
young and $o$ for old. Note the expanded vertical scale compared with Fig.1.
Final populations near $10^7$.
}
\end{figure}

\section{Models}

We start with the standard asexual Penna model program listed in \cite{book} and
discuss here only the changes. The Verhulst deaths, due to the lack of food 
and space in densely populated ecosystems, were restricted to the newborn
\cite{martins} to make the model more realistic; an additional death probability
$\pi = (1 + p)/2)$ at every time step kills individuals of all ages; initially
$p$ is set to an age-independent input parameter $-a$.

At every iteration a new type of inheritable mutations occurs: Randomly $\pi$ is
changed by an amount $+\delta/2$ for young age and by the opposite amount
$-\delta/2$ for old age. The sign of $\delta$ is determined randomly for each
individual at each time step, with equal probabilities for positive and negative
signs. The absolute value $|\delta|$ is fixed as the input parameter $b$. The
summation of the many mutations of $\pi$ leads to a probability difference 
$\Delta = p(t) - p(t=0)$, which is twice the change in the death probability 
$\pi$. Thus a positive $\Delta$ means that the old sacrifice their lifes for the 
young, with a certain probability. (Mutations violating the requirement
$-1 + b < p < 1 - b$ were ignored.)

The minimum reproduction age was taken as 8 age units, and thus ``young'' was
defined as having an age below an input parameter $y$, while old individuals
had an age above $o$; typically, $y = 6, \, o = 10, \, a = 0.8, \, b = 0.02$.
After 2000 iterations the total population as well as the average probability
difference $\Delta$ barely changed anymore, and we stopped the simulation;
in one case this stabilization was confirmed by ten times more iterations. 

This was our first model. A second model assumed the above additional probability 
$p$ not to change with time (no mutations for $p$); instead if a young individual
is supposed to die because of this fixed probability $\pi = (1+a)/2$, it tries to
find an old individual which then dies instead of the young one. The young
individual is allowed up to 20 consecutive attempts to find, by randomly 
selecting one of the existing individuals, one victim old enough to take 
its place on death row.

\section{Results}

Fig.1 shows the emergence of altruism in the sense of the first model: With 
suitable parameters we mostly found the old ones to sacrifice their lives 
for the young, i.e. $\Delta > 0$. However, one of the 
simulations shown there gives negative $\Delta$ where the young die for the old.
In this figure we varied $a$ and $b$ as shown in the headline, and kept 
$y = 6, \, o = 10$ the same. 
Fig.2 shows that varying the limits $y$ and $o$ for young and old age changes
the results much less than varying $a$ and $b$ in Fig.1.

Our second model for the above parameters is bad for altruism; simulations
with and without altruism under otherwise identical conditions showed altruism 
to diminish and not to enhance the population,
at best we could achieve that it did not change the population size for 
long times. 

{\bf Acknowledgements}
We thank the European Science Foundation for supporting the visit of DS
via COST-P10, and Fundacja Na Rzecz Nauki Polskiej for supporting SC.

\end{document}